\documentclass[aps,amsmath,twocolumn,amssymb,titlepage,10pt]{revtex4-1}
\usepackage[T1]{fontenc}
\usepackage[utf8]{inputenc}
\usepackage{amsmath}
\usepackage{braket}
\usepackage{amsfonts}
\usepackage{comment}
\usepackage{graphicx}
\usepackage{hyperref}
\usepackage[capitalize]{cleveref}
\usepackage{amssymb}
\usepackage{amsmath}
\usepackage{mathtools}
\usepackage{todonotes}

\begin{document}
\title{Sublattice-selective percolation on  bipartite planar lattices}
\author{Jonas Wattendorff}
\author{Stefan Wessel}
\affiliation{Institute for Theoretical Solid State Physics, RWTH Aachen University, JARA Fundamentals of Future Information Technology, and \\ JARA Center for Simulation and Data Science, 52056 Aachen, Germany}
\begin{abstract}
In conventional site percolation, all lattice sites are occupied with the same probability. For a bipartite lattice, sublattice-selective percolation instead involves two independent occupation probabilities, depending on the sublattice to which a given site belongs. Here, we determine the corresponding phase diagram for the two-dimensional  square and Lieb lattices from quantifying the parameter regime where a percolating cluster persists for sublattice-selective percolation. For this purpose, we present an adapted Newman--Ziff algorithm. We also consider the critical exponents at the percolation transition, confirming previous Monte Carlo and renormalization-group findings that suggest sublattice-selective percolation to belong to the same universality class as conventional site percolation. To further strengthen this conclusion, we  finally treat sublattice-selective percolation on  the Bethe lattice (infinite Cayley tree) by an exact solution. 
\end{abstract}
\maketitle

\section{Introduction}\label{Sec:Introduction}

Percolation provides a remarkably simple route to nontrivial critical phenomena, which have a broad range of applications in many fields of science and engineering~\cite{Flory1941, Stauffer2018}. In its most basic formulation, one considers an infinite periodic lattice, occupying each lattice site independently with equal probability $p$. The occupied sites form contiguous clusters of connected lattice sites, which turn out to exhibit several interesting properties. In particular, an infinite, spanning (percolating) cluster of connected occupied sites emerges for sufficiently large values of $p$ beyond a particular threshold value $p_c$. For example, considering a two-dimensional square lattice, the site percolation threshold $p^\mathrm{sq}_c=0.592 746...$~\cite{Lee2008, Jacobsen2015} is known from computational studies to rather high accuracy --  the exact value being, however, unknown to date. In the vicinity of the percolation threshold, various characteristics of the cluster distribution exhibit non-trivial scaling behavior: For $p \gtrsim p_c$, the strength $P$ of the percolating cluster, i.e., the  probability that a given site belongs to the percolating cluster, increases with $p$ by an asymptotic power law,
\begin{equation}
P\propto (p-p_c)^\beta,
\end{equation}
resembling the scaling behavior of an order parameter near a thermal ordering phase transition. Upon approaching $p_c$, the correlation length $\xi$, defined as the mean distance between two sites belonging to the same finite cluster, increases as 
\begin{equation}
\xi \propto |p-p_c|^{-\nu},
\end{equation}
and the mean number of sites of a finite cluster $S$ increases as
\begin{equation}
S \propto |p-p_c|^{-\gamma},
\end{equation}
corresponding to the order parameter susceptibility in thermal ordering transitions. The scaling exponents describe the critical behavior at the percolation transition of the above state functions, and they are considered  universal in the sense that within a given dimension $d$ they do not depend on details of the lattice structure (e.g., square or triangular in $d=2$) or the kind of percolation problem considered (site, bond, or also in the continuum)~\cite{Stauffer2018}.
For $d=2$ the exponents are known as $\beta=5/36=0.13\bar{8}$, $\nu=4/3$, and $\gamma=43/18=2.3\bar{8}$, respectively~\cite{Smirnov2001, Stauffer2018}. Furthermore, at the percolation threshold, the infinite cluster has a fractal dimension $d_f=d-\beta/\nu$, e.g.,   $d_f=91/48=1.8958\bar{3}$ for the two-dimensional case. 

\begin{figure}[t]
    \centering
    \includegraphics[width=0.45\textwidth]{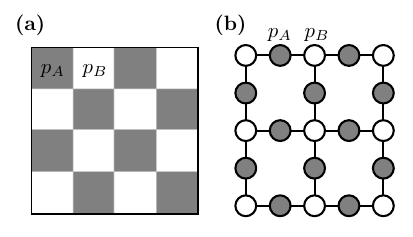}
    \caption{Illustration of the square (a) and Lieb lattice (b). Sites from sublattice $A$ and $B$ are shown in grey (white). The square lattice is shown as a checkerboard, while the nearest-neighbor bonds of the Lieb lattice are shown explicitly.}
    \label{Fig:lattices}
\end{figure}

In physical systems the lattice sites considered above may represent objects such as atoms in a solid state crystal, an empty site being related to a defect, e.g., an impurity atom. In many condensed-matter systems, the underlying lattice structure shows further characteristics, such as being bipartite, i.e., the total set of lattice sites can be decomposed into two disjoint sublattices, denoted $A$ and $B$ in the following, with nearest-neighbor sites always belonging to different sublattices (cf. Fig.~\ref{Fig:lattices} for two examples). If, for example in a binary system, the objects occupying the $A$ and $B$ sublattice are of different species, it becomes feasible that the susceptibility to a defect differs on the $A$ and $B$ sublattice. In the context of site percolation, such a circumstance can be described by assigning two different occupation probabilities $p_A$ and $p_B$ to sites from sublattices $A$ and $B$, respectively. Such sublattice-selective percolation was introduced by Scholl and Binder in Ref.~\cite{Scholl1980}, motivated by the  properties of diluted magnetic compounds, with a special focus towards experimental findings in three-dimensional spinel structures. For the latter case, they performed Monte Carlo simulations to extract the phase diagram in terms of percolating vs. non-percolating regimes of parameters $(p_A,p_B)$, and they provided numerical evidence that the critical exponents of sublattice-selective percolation are those of the universality class for conventional site percolation (where $p_A=p_B$). Reference \cite{Scholl1980} also considered several other lattice structures, both in three and two dimensions, focusing on the extreme case in which one sublattice is fully occupied, e.g., $p_B=1$. In various cases, the critical value $p_{A,c}$ for $p_A$ can then be expressed in terms of the percolation threshold of certain related conventional percolation problems. For example, for a square lattice with $p_B=1$, the threshold value $p_{A,c}$ equals the site percolation threshold for a square lattice with next-nearest neighbor connectivity. Using the concept of matching lattices, this value can furthermore be shown to be equal to $1-p^\mathrm{sq}_c=0.407253...$~\cite{Sykes1964}. For $p_B=1$, the percolation threshold for $p_A$ thus falls below the value of $p^\mathrm{sq}_c$ -- as expected, since half of the lattice is already occupied for $p_B=1$. Due to the equivalence between the two sublattices of the square lattice, the problem is symmetric under the exchange of $p_A$ and $p_B$ in this case. In particular, $p_{B,c}=1-p^\mathrm{sq}_c$ along the line $p_A=1$, and along the symmetric line $p_A=p_B$, the critical value  $p^\mathrm{sq}_c$ is recovered. A quantification of the full boundary line in the $(p_A, p_B)$-plane of the percolating regime was however not performed in  Ref.~\cite{Scholl1980} for the basic square lattice case. Later, approximate renormalization-group (RG) calculations were used to estimate this boundary line~\cite{Idogaki1982a} and to address the question of universality (the method was furthermore employed to study sublattice-selective percolation on other lattices and sublattice geometries~\cite{Idogaki1982b, Idogaki1983a}). 
However, while the RG approach finds that sublattice-selective percolation is indeed controlled by the same fixed point as the uniform case, its numerical accuracy is limited. For example, in the uniform case a best estimate of 0.610  was obtained for $p^\mathrm{sq}_c$~\cite{Idogaki1982a}. Related work, using a decoupling approximation for an effective field theory of the Ising model with sublattice-selective depletion yields results of a similar accuracy~\cite{Idogaki1995,Ueda2000}. Since those early works on sublattice-selective percolation, several methodological  advances have been achieved, such as the Newman--Ziff algorithm~\cite{Newman2000,Newman2001}, which allows for efficient high-accuracy numerical studies of conventional lattice-based percolation problems. 
Recentlty it has been shown~\cite{rosales_herrera_percolation-intercropping_2021} that it is now feasible to also perform a systematic and accurate exploration of sublattice-selective percolation on basic planar lattices, such as the square lattice using such advanced algorithms.

Here, we perform such an analysis using an adapted version of the Newman--Ziff algorithm~\cite{Newman2000,Newman2001}, which we detail further below. This algorithm allows for efficient Monte Carlo studies of the phase diagram, like the original algorithm does for conventional percolation problems on periodic lattices.  In addition to the square lattice, we  examine the case of the planar Lieb lattice, which is also bipartite, and for which conventional site-percolation has been considered recently~\cite{Oliveria2021}. This lattice, which is of interest in the context of flat band physics, ferrimagnetism, and topological states, is a decorated square lattice and bipartite; cf.~Fig.~\ref{Fig:lattices}. However, in contrast to the square lattice, the two sublattices $A$ and $B$ of the Lieb lattice are not equivalent, and the phase diagram of sublattice-selective percolation is non-symmetric in the $(p_A,p_B)$-plane. Furthermore, we use finite-size scaling in order to estimate the critical exponents for sublattice-selective percolation on the square lattice. Our results support previous conclusions~\cite{Scholl1980, Idogaki1982a} regarding the universal properties. 

\begin{figure}[t]
    \centering
    \includegraphics[trim={0cm 1cm 0cm 0.5cm}, width=0.35
    \textwidth]{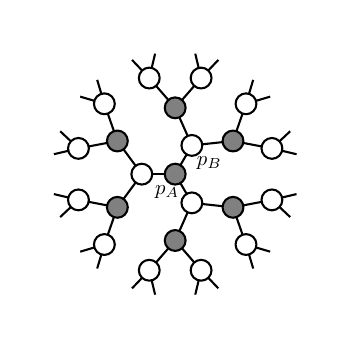}
    \caption{Illustration of the Bethe lattice for $z=3$. Sites from sublattices $A$ and $B$ are shown in gray (white). Note that for the infinite lattice, all sites are in fact topologically equivalent.}
    \label{Fig:Bethe}
\end{figure}

The remainder of this article is organized as follows: In Sec.~\ref{Sec:NewmanZiff}, we introduce an adapted Newman--Ziff algorithm for efficient computational studies of sublattice-selective percolation. Then, we report our results for the square and Lieb lattice in Secs.~\ref{Sec:Square} and ~\ref{Sec:Lieb}, respectively. 
Finally, in Sec.~\ref{Sec:Bethe}, we study sublattice-selective percolation on the Bethe lattice, which in various aspects corresponds to infinite dimension, $d=\infty$, for coordination numbers $z\geq 3 $, cf. Fig.~\ref{Fig:Bethe} for an illustration for $z=3$. We determine the critical exponents analytically for the cases $z=2$ (corresponding to the chain, $d=1$) and $z=3$, thereby demonstrating  explicitly the anticipated universality. Our final conclusions are given in Sec.~\ref{Sec:Conclusions}.

\section{Bipartite Newman--Ziff Algorithm}\label{Sec:NewmanZiff}

The Newman--Ziff algorithm~\cite{Newman2000,Newman2001} performs Monte Carlo sampling of the percolation problem on finite lattices of linear size $L$ and with $N\propto L^2$ lattice sites in two dimensions. For a given system size, one measures appropriate finite-size estimators $Q_L$ for any of the state functions $Q$, such as the quantities $P$, $\xi$, and $S$ introduced above. In the following, we consider finite systems with periodic boundaries, and a cluster is defined to be percolating if it completely wraps around in either direction.  
The main idea behind the Newman--Ziff algorithm can be loosely described as first calculating a given state function $Q_L$ in the microcanonical ensemble and then transforming to the canonical ensemble. In the microcanonical ensemble for a bipartite lattice, the state function is a function $Q_L(n_A, n_B)$ of the number of occupied sites $n_A$ and $n_B$ for each sublattice, while in the canonical ensemble it is a function of the probabilities, $Q_L(p_A, p_B)$. This is an analogy to thermodynamics, where $n_A$ corresponds to the energy of the system and $p_A$ corresponds to the temperature.

To explore the phase diagram, we take cuts for fixed values of $p_B$, which means populating each site on the $B$ sublattice with probability $p_B$ and then successively and randomly adding sites to the $A$ sublattice until it is full, thus calculating $Q_L(n_A, p_B)$ for all values of $n_A$. Then the transformation to the canonical ensemble, i.e., $Q_L(n_A, p_B)\rightarrow Q_L(p_A, p_B)$, is performed for the $A$ sublattice only. For a given value of $p_A$, the probability $B(N_A, n_A, p_A)$ that there are exactly $n_A$ sites occupied from the total number of $N_A$ sites of the $A$ sublattice is given by
\begin{equation}
    B(N_A, n_A, p_A)=\binom{N_A}{n_A}p_A^{n_A}(1-p_A)^{N_A-n_A},\label{eq:NZ_binom}
\end{equation}
since choosing the occupied sites is a Bernoulli process. Weighing the (with respect to the $A$ sublattice) microcanonical state function $Q_L(n_A, p_B)$  by the probability in Eq.~(\ref{eq:NZ_binom}) and summing over all values of $n_A$ gives the canonical state function $Q_L(p_A, p_B)$,
\begin{equation}
    Q_L(p_A, p_B)=\sum_{n_A=0}^{N_A}B(N_A, n_A, p_A)Q_L(n_A, p_B),\label{eq:NZ_canon_trans}
\end{equation}
which is essentially a convolution with the binomial distribution. This transformation is again analogous to thermodynamics, where the transformation is performed via the Boltzmann distribution instead of the binomial distribution.

The efficiency of the Newman--Ziff algorithm derives from the fact that the number of samples for $p_A$ is only limited by the cost of the transformation in Eq.~(\ref{eq:NZ_canon_trans}) and that this transformation is linear in the state function $Q_L$. Hence, only the average of the microcanonical state function after many Monte Carlo steps has to be transformed explicitly,
\begin{equation}
    \langle Q_L(p_A, p_B)\rangle=\sum_{n_A=0}^{N_A}B(N_A, n_A, p_A)\langle Q_L(n_A, p_B)\rangle.\label{eq:NZ_canon_trans_mean}
\end{equation}

\begin{figure}[t]
    \centering
    \includegraphics[width=0.25\textwidth]{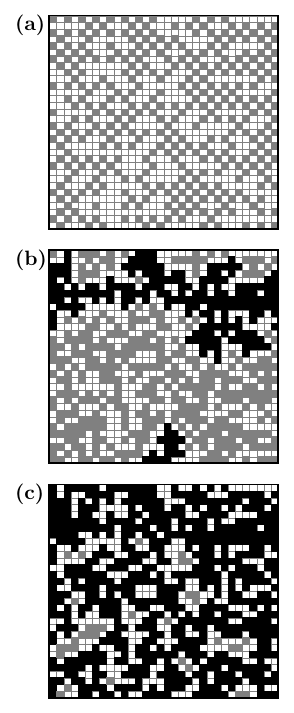}
    \caption{Illustration of the sublattice-selective percolation process  for an $L=32$ square lattice and periodic boundary conditions. Occupied (empty) sites are shown in gray (white), and the percolating cluster is in black. Starting from a random occupation of the $B$ sublattice sites with probability $p_B=0.7$, the $A$ sublattice is empty in (a), filled by $n_A=225$ sites in (b), and $n_A=300$ sites in (c).}
    \label{Fig:NZ}
\end{figure}

The algorithm thus consists of three main steps. Firstly, every site on the $B$ sublattice is occupied with a given probability $p_B$, as seen in the example of Fig.~\ref{Fig:NZ}(a). Since the lattice is bipartite, there are no clusters of size greater that single sites at this point. 

Secondly, sites on the $A$ sublattice are randomly occupied according to a random permutation of the sublattice indices. Every time a site is occupied, one checks each nearest neighbor one by one, and if the neighbor is occupied and belongs to a different cluster these are merged using a union--find routine, as in the original Newman--Ziff algorithm. Every time a site is newly occupied, the state functions $Q_L(n_A, p_B)$ are added to the mean $\langle Q_L(n_A, p_B)\rangle$. In Fig.~\ref{Fig:NZ}(b), the state in which the cluster first percolates (with periodic boundary conditions) is shown. In Fig.~\ref{Fig:NZ}(c) , the lattice is shown after adding 75 occupied sites, and only a few small clusters besides the percolating cluster remain.

Thirdly, all the state functions are transformed according 
to Eq.~(\ref{eq:NZ_canon_trans_mean}). The binomial coefficients are computed recursively~\cite{Newman2001}, as
\begin{equation}
    B(N_A, n_A, p_A)=
    \begin{dcases}
        B(N_A, n_A-1, p_A) \frac{N_A-n_A+1}{n_A}\frac{p_A}{\bar p_A}\\
        B(N_A, n_A+1, p_A) \frac{n_A+1}{N_A-n_A}\frac{\bar p_A}{p_A}
    \end{dcases},\nonumber
\end{equation}
for $n_A>p_A N_A$ and $n_A<p_A N_A$, respectively, with  $\bar{p}_A=1-p_A$.
In practice, $B(N_A, n_A, p_A)$ is negligible for many values of $n_A$, so only seven standard deviations $\sqrt{p_A(1-p_A)N_A}$ around the maximum value $ p_A N_A$ are actually calculated, which only excludes $B(N_A, n_A, p_A)\lesssim 10^{-10}$.

\begin{figure}[t]
     \centering
      \includegraphics[width=0.25\textwidth]{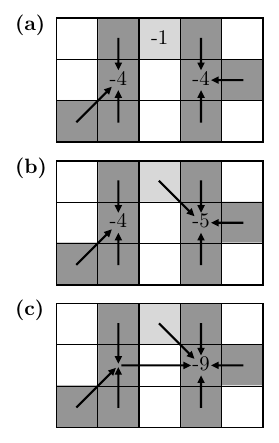}
     \caption{Example of adding a newly occupied site on a square lattice. Gray sites are occupied, and the light gray site is newly occupied. The roots are marked by negative numbers, and the arrows symbolize pointers.}
     \label{Fig:NZ_example}
\end{figure}

In the actual implementation, the lattice is represented by an integer array of size $N$. If a site is occupied, it is part of a tree, and each tree corresponds to one cluster. A cluster of size $s$ has one root, which has an array entry of value $-s$, while the other sites of the cluster have pointers (indices of the array) which eventually lead to the root. 
The idea behind the aforementioned union--find algorithm is that the union operation merges two trees and the find operation returns the root of a given occupied site. So if there are two neighboring occupied sites on the lattice, the roots can be compared via find, and if the roots are different, the trees will be merged via union. Figure~\ref{Fig:NZ_example} shows an example of this procedure. The light gray site in Fig.~\ref{Fig:NZ_example}(a) has been newly added, making it a tree just consisting of a root. Next, it is merged with the right cluster in Fig.~\ref{Fig:NZ_example}(b), making a larger cluster of size $s=5$. In the next step in  Fig.~\ref{Fig:NZ_example}(c), the roots of the light gray site and the site to the left of it are compared via the find operation, and the roots are at different lattice sites. Thus, the union operation is invoked, and the root of the left tree then points to the root of the right tree and the cluster sizes are added. Now every site in the cluster returns the same root with find.
There are a couple of methods to make this procedure more efficient: First, the so-called path compression is the idea that the find operation is most efficient when the path to the root involves as few pointers as possible. So every time the find operation follows a path from some site to the root, all the pointers on its way are set to point directly to the root -- in practice, this makes the trees never deeper than a few generations.
Secondly, a smaller cluster is always appended to a bigger one. The find operation for the cluster that is appended takes one step longer, hence by appending the smaller cluster, the average number of steps is again minimized.

In terms of computational complexity, using approaches based on alternative methods, such as the Hoshen--Kopelman algorithm~\cite{Hoshen1976}, one would have to build all clusters for every value of $p_A$ from scratch, and the time complexity would be $\mathcal{O}(N_A^2)$, for each realization for a given value of $p_B$.
In contrast, the Newman--Ziff algorithm takes time of order $\mathcal{O}(N_A+M_T)$ to calculate the $N_A$ samples of $Q_L(n_A, p_B)$ and transform from the microcanonical to the canonical ensemble, where $M_T$ is the number of steps needed to calculate the sum and binomial distribution in the transformation of Eq.~(\ref{eq:NZ_canon_trans_mean}). 
For the system sizes used here, $M_T$ and thus the computational time for the transformation are comparably small. For a detailed discussion of the computational advantage of the  Newman--Ziff algorithm, we refer the reader to Ref.~\cite{Newman2001}.

\section{Square lattice}\label{Sec:Square}
We used the  bipartite Newman--Ziff algorithm algorithm to examine  sublattice-selective percolation on the square lattice, and we report our numerical results in this section. 
\subsection{Percolation threshold}

\begin{figure}[t]
     \centering
      \includegraphics[width=0.45\textwidth]{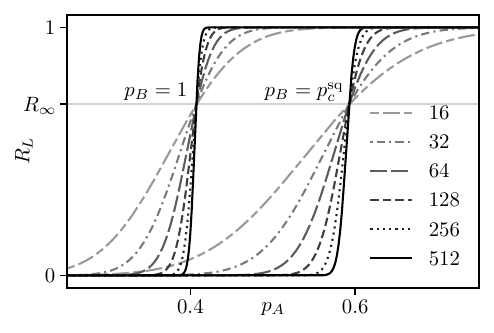}
     \caption{Numerical results for the dependence of $R_L$ as a function of $p_A$ for different values of $L$ for sublattice-selective percolation on the square lattice for  $p_B=1$ (left curves) and  $p_B=p_c^\mathrm{sq}$ (right curves). The horizontal line indicates the exact value of $R_\infty$ at $p_c^\mathrm{sq}$ from Ref.~\cite{Newman2001}.}
     \label{Fig:sq_RL}
\end{figure}

We first consider the determination of the percolation threshold line for sublattice-selective percolation on the square lattice. As noted in the previous section, for this purpose we consider a set of fixed values of $p_B$, and then we use the bipartite Newman-Ziff algorithm to obtain $Q_L(p_A,p_B)$ for essentially any value of $p_A$. Denoting by $R_L$ the probability for the existence of a percolating cluster (and estimated by the Monte Carlo mean), this state function increases monotonously from 0 to 1 smoothly upon increasing $p_A$ for finite $L$. This behavior is shown for two different values of $p_B$ in Fig.~\ref{Fig:sq_RL}, considering $p_B=p_c^\mathrm{sq}$, and $p_B=1$, respectively. 

In the infinite system limit, this quantity -- denoted $R_\infty(p_A)$ -- exhibits a jump at the percolation threshold and we can obtain an estimate $p_{A,c}(L)$ of $p_{A,c}$ at the specified value of $p_B$ from the condition $R_L(p_{A,c}(L))=C$, where $C\in(0,1)$ is kept fixed upon varying $L$. The speed of convergence of $p_{A,c}(L)$ towards $p_{A,c}$ varies for different choices of $C$. For conventional site percolation, Newman and Ziff~\cite{Newman2001} suggest taking $C$ equal to the exactly known value of $R_\infty$ at $p_c^\mathrm{sq}$, which equals $0.6904...$, and finding an algebraic asymptotic convergence, 
\begin{equation}\label{Eq:qc_fs}
p_{A,c}(L)-p_{A,c} \propto L^{-X},
\end{equation}
conjecturing that $X=2+1/\nu$, which equals $2.75$ for $d=2$, leading thus to fast convergence. 
From our simulations, we observe no significant change in  the threshold value of  $R_\infty$ for the sublattice-selective case (cf., e.g.,  the crossing point values in Fig.~\ref{Fig:sq_RL}). A formal proof of this statement would, however, still be valuable. In any case, we always fixed $C$ to the above value in order to fit $p_{A,c}(L)$ to the finite-size scaling in Eq.~(\ref{Eq:qc_fs}). 

In particular, we performed finite-size calculations for system sizes between $L=16$ and $L=512$, doubling consecutive values of $L$. For each fixed value of $p_B$, we scaled the number of Monte Carlo steps (each corresponding to an initial configurations as in Fig.~\ref{Fig:NZ}(a), followed by successively increasing $n_A$) with $L^{-3/2}$, taking $10^7$ steps at $L=16$. This procedure ensures a similar absolute statistical uncertainty on the estimator for $R_L(p_{A,c}(L))$ upon varying $L$,  as it does for conventional percolation~\cite{Newman2001}: For a given probability $p_A$, the state function $R_L(p_A)$ is either 1 or 0, so the uncertainty comes from the binomial distribution,
\begin{equation}
    \sigma_{R_L}=\sqrt{\frac{R_L(p_A)(1-R_L(p_A))}{M}},\label{eq:NZ_R_err}
\end{equation}
where $M$ is the number of Monte Carlo steps. As the width of the critical region decreases as $L^{-1/\nu}$, the gradient $dR_L/dp_A$ of $R_L(p_A)$ in the critical region increases as $L^{1/\nu}$. Together with Eq.~(\ref{eq:NZ_R_err}), the uncertainty on $p_{A,c}$ thus scales as $\sigma_{p_{A,c}}\sim L^{-1/\nu}\sigma_{R_L}\sim L^{-1/\nu}M^{-1/2}$. To keep $\sigma_{p_{A,c}}$ constant, $M$ thus has to scale with $L^{-2/\nu}$. For $d=2$, the universal value is $\nu=4/3$, and this value will also be used for this uncertainty calculation here (this has no impact on the estimation of the critical exponents as performed in the next section -- there will indeed also be no evidence that $\nu\neq 4/3$).
Overall, the whole procedure was repeated 100 times for a given $p_B$ in order to obtain the statistical error on all other estimated state functions. For the final transformation from $n_A$ to $p_A$, we used a narrow $p_A$-grid with a spacing of order $10^{-4}$.

\begin{figure}[t]
     \centering
      \includegraphics[width=0.35\textwidth]{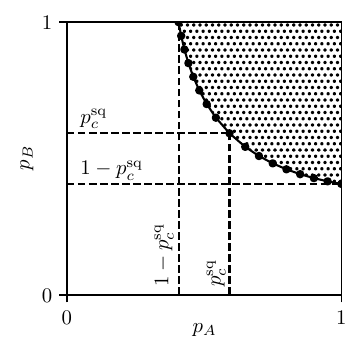}
     \caption{Phase diagram for sublattice-selective percolation on the square lattice, as obtained from the bipartite Newman--Ziff algorithm. The dotted regime is percolating.}
     \label{Fig:sq_phasediag}
\end{figure}

Based on the numerical data, we obtain the values of $p_{A,c}$ for several values of $p_B$ reported in Table~\ref{Tab:sq}. Also included in this table are the values of the exponent $X$ that we obtain based on these fits (the values of $\chi^2/\mathrm{d.o.f.}$ (degree of freedom) are of $O(1)$). Overall, the estimated values of $X$ exhibit some scatter, but no significant systematic deviation from the proposed value for the conventional case~\cite{Newman2001}. The numerical results for $p_{A,c}$ that we obtain for the specific cases of $p_B=p_c^\mathrm{sq}$ and $p_B=1$ are in accord with the previously reported values, ensuring the overall accuracy of our approach. 
\begin{table}[t]
    \centering
    \begin{tabular}{|l||l|l|l|}
    \hline
        $p_B$      & $p_{A,c}$          & $X$            & $\nu$  \\ \hline \hline
        $p_c^\mathrm{sq}$ & $0.5927483(16)$    & $2.80(10)$     &  $1.3318(4)$    \\ \hline
        $0.65$     & $0.5431265(18)$    & $2.66(13)$     &  $1.331(4)$   \\ \hline
        $0.7$      & $0.5095937(13)$    & $2.49(7)$      &  $1.3290(7)$   \\ \hline
        $0.75$     & $0.4826738(13)$    & $2.42(7)$      &  $1.3340(6)$  \\ \hline
        $0.8$      & $0.46082758(35)$   & $2.561(18)$    &  $1.3352(10)$    \\ \hline
        $0.85$     & $0.4430110(13)$    & $2.60(6)$      &  $1.3368(12)$    \\ \hline
        $0.9$      & $0.4284801(17)$    & $2.56(8)$      &  $1.33241(28)$    \\ \hline
        $0.95$     & $0.4166941(8)$     & $2.598(34)$    &  $1.3321(4)$  \\ \hline
        $1$        & $0.4072524(6)$     & $2.556(21)$    &  $1.3337(5)$\\
         \hline
    \end{tabular}
    \caption{Numerical results obtained for sublattice-selective percolation on the square lattice using the bipartite Newman--Ziff algorithm.}
    \label{Tab:sq}
\end{table}
Based on the numerical values of $p_{A,c}$, and using the symmetry under the exchange of $p_A$ and $p_B$, we obtain the phase diagram for sublattice-selective percolation on the square lattice shown in Fig.~\ref{Fig:sq_phasediag}.

\subsection{Critical exponents}

\begin{figure}[t]
     \centering
      \includegraphics[width=0.35\textwidth]{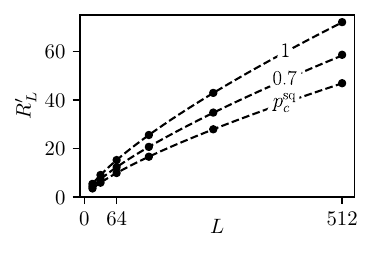}
     \caption{Fits of the numerical data for $R'_L$ to the finite-size scaling form in Eq.~(\ref{Eq:RpL_scaling}) for different indicated values of $p_B$.}
     \label{Fig:sq_RpL}
\end{figure}

\begin{figure}[t]
     \centering
      \includegraphics[width=0.5\textwidth]{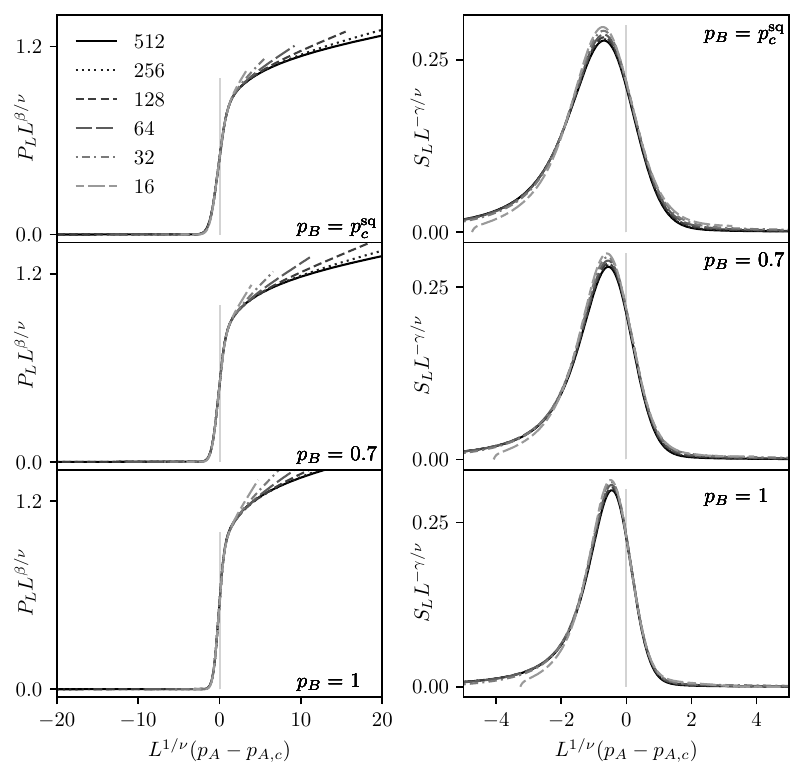}
     \caption{Data collapse plots for finite-size data labeled by $L$ for $P$ and $S$ for different values of $p_B$ using the critical exponents for $d=2$ percolation. }
     \label{Fig:sq_PS}
\end{figure}

In addition to the percolation threshold, we  estimated the critical exponents for sublattice-selective percolation on the square lattice using appropriate finite-size scaling analysis~\cite{Stauffer2018}. We first consider the critical exponent $\nu$, which can best be estimated from the finite-size scaling of the gradient $R'_L=dR_L/dp_A$ of $R_L(p_A)$ near the percolation threshold,
\begin{equation}\label{Eq:RpL_scaling}
R'_L \propto L^{1/\nu}.
\end{equation}
We approximate $R'_L$ from the slope of  a linear regression within a $p_A$ range of $10^{-3}$ around the estimated value of $p_{A,c}$. The thus obtained estimates for $\nu$ are given in Table~\ref{Tab:sq}, and Fig.~\ref{Fig:sq_RpL} shows bests fits to the numerical data for different values of $p_B$. 
The values that we obtain for $\nu$ scatter with no systematic dependence of $p_B$, and taking their average gives $\nu=1.3352(11)$, which is in good agreement with $\nu=4/3$ for the $d=2$ universality class. We also find no indication that the critical exponents $\beta$ and $\gamma$ differ from their conventional values for $d=2$. To access these exponents, we consider the finite-size scaling forms
\begin{equation}
    P_L=L^{-\beta/\nu} F_P[L^{1/\nu} (p_A-p_{A,c})]
\end{equation}
for the estimate $P_L$ of the strength of the percolating cluster in terms of a scaling function $F_P$, and similarly
\begin{equation}
    S_L=L^{\gamma/\nu} F_S[L^{1/\nu} (p_A-p_{A,c})]
\end{equation}
for the mean number $S$ of finite clusters. From appropriate data-collapse plots, one can adjust the values of $\beta$ and $\gamma$, based on our previous estimates for $\nu$ and $p_{A,c}$, in order to obtain the best collapse within the critical region. Using such an analysis, we find that the best-fit values are in fact in accord with the known values for conventional percolation in $d=2$. This is illustrated for several values of $p_B$ in Fig.~\ref{Fig:sq_PS}. We thus conclude from our finite-size analysis, that the critical exponents for sublattice-selective percolation are in agreement with the universal values for $d=2$ percolation, supporting previous findings~\cite{Scholl1980, Idogaki1982a}. 

\section{Lieb lattice}\label{Sec:Lieb}

\begin{figure}[t]
     \centering
     \includegraphics[width=0.35\textwidth]{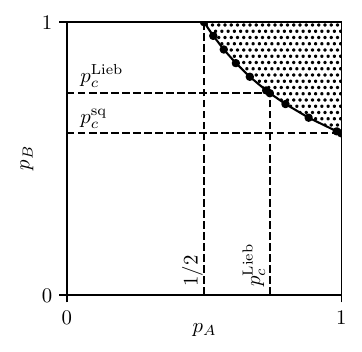}
     \caption{Phase diagram for sublattice-selective percolation on the Lieb lattice, as obtained from the bipartite Newman--Ziff algorithm. 
     The dotted regime is percolating.}
     \label{Fig:Lieb}
\end{figure}

As a further example, we examine sublattice-selective percolation on the Lieb lattice, for which the case of uniform site percolation ($p_A=p_B$) has recently been considered~\cite{Oliveria2021}. Since the Lieb lattice is in fact a decorated square lattice, sublattice-selective percolation is equivalent in this case to mixed bond-site percolation on the square lattice, with $p_A$ ($p_B$) specifying the bond (site) occupation probability. In particular, in the limit $p_A=1$ ($p_B=1)$, we obtain conventional pure site (bond) percolation on the square lattice, and thus $p_{B,c}=p_c^\mathrm{sq}$ ($p_{A,c}=1/2$) along these lines, respectively (here, we used the fact that the percolation threshold $1/2$ for bond percolation on the square lattice has been exactly determined~\cite{Kesten1980}). This shows explicitly that due to the inequivalence of the two sublattices of the Lieb lattice, the phase diagram is not symmetric in the $(p_A,p_B)$-plane. 
In Ref.~\cite{Tarasevich1999}, the general case of mixed bond-site percolation is considered for various lattices, including the square lattice (mixed bond-site percolation on several lattices is also considered in Refs.~\cite{Yanuka1990,Gonzalez2013,Gonzalez2021,Torres2022}) . The phase diagram was numerically estimated, but only to comparably low accuracy, so that here we employed the bipartite Newman-Ziff algorithm in order to obtain more accurate values for the threshold line for sublattice-selective percolation on the Lieb lattice. The numerical results are provided in Table~\ref{Tab:Lieb}, and the corresponding phase diagram is shown in Fig.~\ref{Fig:Lieb}. Comparing the value of $p_{A,c}$ obtained for $p_B=1$ to the exact value ($1/2$) confirms the accuracy of our results. For estimating these numbers, $C$ was chosen identical to the square lattice value from the previous section, resulting in similarly fast convergence.
Additionally, the previous estimate~\cite{Oliveria2021} for the conventional ($p_A=p_B$) site percolation threshold $p_c^{\mathrm{Lieb}}$ on the Lieb lattice was improved using the Newman--Ziff algorithm. Along this line, we find that $C=0.68$ yields better convergence (however, even then the exponent $X\approx 1/\nu$ is significantly lower). The obtained value  $p_c^{\mathrm{Lieb}}=0.7397120(17)$ is consistent with the previous best estimate~\cite{Oliveria2021}.
\begin{table}[t]
    \centering
    \begin{tabular}{|l||l|l|l|}
    \hline
        $p_B$      & $p_{A,c}$          \\ \hline \hline
        $0.6$      & $0.983336(4)$        \\ \hline
        $0.65$     & $0.8809405(10)$     \\ \hline
        $0.7$      & $0.7965947(21)$       \\ \hline
        $0.75$     & $0.7261839(20)$     \\ \hline
        $0.8$      & $0.6666702(16)$      \\ \hline
        $0.85$     & $0.6157944(18)$        \\ \hline
        $0.9$      & $0.5718775(14)$      \\ \hline
        $0.95$     & $0.5336113(21)$      \\ \hline
        $1$        & $0.4999987(19)$     \\
         \hline
    \end{tabular}
    \caption{Numerical results obtained for sublattice-selective percolation on the Lieb lattice using the bipartite Newman--Ziff algorithm.}
    \label{Tab:Lieb}
\end{table}

\section{Bethe lattice}\label{Sec:Bethe}

To complement the above numerical study by exact analytical results, we finally consider sublattice-selective percolation on the Bethe lattice. We first consider the case $z=2$, essentially corresponding to the one-dimensional chain.

\subsection{Percolation on a chain ($z=2$)}

In this case, percolation occurs only at the singular point $(p_A,p_B)=(1,1)$. To extract the critical exponents upon approaching the percolation point along different lines within the $(p_A,p_B)$-plane, we next examine appropriate state functions. 

A cluster containing $s$ occupied sites with one empty site to either side is called an $s$--cluster. The number of $s$--clusters per lattice site $n_s$ is a useful state function. Another way of looking at $n_s$ is that it is the probability of an arbitrary site to be the left-end site of an $s$--cluster. This is the case if the site to the left is empty, then $s$ consecutive sites are occupied, followed by another empty site. For $p_A=p_B=p$, the state function would be $n_s=(1-p)p^2(1-p)$, but for $p_A\neq p_B$ one has to consider whether the left end site is from the $A$ or $B$ sublattice, and whether $s$ is even or odd. The probability that an arbitrary site is from the $A$ or $B$ sublattice is equal to $1/2$, and from there on it is a matter of counting the $A$-- and $B$--sublattice sites in the cluster to find
\begin{equation}
    n_s(p_A, p_B)=
        \frac12 \left((\bar p_B)^2p_A+(\bar p_A)^2p_B\right)p_A^{(s-1)/2}p_B^{(s-1)/2}
\end{equation}
for odd $s$, and
\begin{equation}
    n_s(p_A, p_B)=
        \bar p_A \bar p_B p_A^{s/2}p_B^{s/2}
\end{equation}
for even $s$, where $\bar{p}_A=1-p_A$, $\bar{p}_B=1-p_B$.
The probability that an arbitrary site is part of an $s$--cluster is $n_s\cdot s$, so the probability that an arbitrary site is part of any cluster is given by
\begin{equation}
    \sum_{s=1}^\infty n_s(p_A, p_B) s =\frac12(p_A+p_B)\label{eq:1d_unity_fin},
\end{equation}
i.e., equivalent to the probability that an arbitrary site is occupied. From $n_s$, the mean cluster size $S$ can be evaluated, which equals the ratio of the second divided by the first moment of the cluster size distribution,
\begin{align}
    S(p_A, p_B) &= \frac{\sum_{s=1}^\infty n_s(p_A, p_B) s^2}{\sum_{s=1}^\infty n_s(p_A, p_B) s}\nonumber\\
    &=\frac{p_A^2p_B+p_Ap_B^2+4p_Ap_B+p_A+p_B}{p_A+p_B-p_A^2p_B-p_Ap_B^2},\label{eq:1d_S_fin}
\end{align}
which approaches 1 for $(p_A,p_B)\rightarrow (0,0)$ and diverges upon approaching $(1,1)$. We obtain the critical exponent $\gamma$ from analyzing the singular behavior of $S$ near $(1,1)$. Fixing $p_B=1$, the asymptotic  singularity is obtained as
\begin{align}
    S(p_A, 1) &= \frac{p_A^2+6p_A+1}{1-p_A^2}\nonumber\\
    &\sim 4(1-p_A)^{-1}, \quad \text{for $p_A\sim 1$},
\end{align}
so that $\gamma =1$. This asymptotic scaling also results upon approaching $(1,1)$ along any other line. Namely, for $p_A=1-r \cos(\phi), p_B=1-r\sin(\phi)$, with $\phi\in[0,\pi/2]$, the asymptotic behavior $S\sim 4/[(\cos(\phi)+\sin(\phi))\: r]$ for $r\rightarrow 0$ is obtained.

To obtain the correlation length $\xi$, we first calculate the correlation function (pair connectivity function)   $g(r)$, which equals the mean number of sites that are in the same cluster as a given occupied site, at a distance $r$ apart. Thus, there have to be $r-1$ consecutive occupied sites in between. For $p_A=p_B=p$ this is just $2p^r$ (2 for both directions), but for $p_A\neq p_B$ the parity of $r$ and the starting site have to be accounted for. The starting site is on the $A$ sublattice with probability $\frac{p_A}{p_A+p_B}$ and on the $B$ sublattice with probability $\frac{p_B}{p_A+p_B}$. Accounting for the parity of $r$ and the special case $r=0$, we obtain
\begin{equation}
    g(r)=
    \begin{dcases}
        1, &r=0,\\
        \frac{4}{p_A+p_B}p_A^{(r+1)/2}p_B^{(r+1)/2}, &\text{$r$ odd},\\
        2p_A^{r/2}p_B^{r/2}, &\text{$r$ even}.
    \end{dcases}
\end{equation}
The correlation length $\xi$ quantifies the length scale of the correlation function, and can also be viewed as the radius of the clusters that give the main contribution to the mean cluster size $S$~\cite{Stauffer2018}. The formal definition is
\begin{equation}
    \xi^2=\frac{\sum_{r} r^2g(r)}{\sum_{r} g(r)}.\label{eq:1d_xi_def}
\end{equation}
A well-known identity in conventional percolation is  that the sum over all $g(r)$  equals  the mean cluster size, and this indeed also follows for sublattice-selective percolation, as we obtain
\begin{equation}
    \sum_{r=0}^{\infty} g(r)=S. 
\end{equation}

The numerator in Eq.~(\ref{eq:1d_xi_def}) can also be calculated, and for $p_B=1$ it scales as $(1-p_A)^{-3}$ near the percolation threshold. Since $1/S$ scales as $(1-p_A)$, the singularity of $\xi$ is thus obtained as
\begin{equation}
    \xi^2(p_A, 1)=\frac{\sum_{r} r^2g(r)}{S}\sim (1-p_A)^{-2}, \quad \text{for $p_A\sim 1$},
\end{equation}
and we extract the critical exponent $\nu=1$ for the divergence of $\xi$ near the percolation threshold. 

The third state function that we consider is the strength of the percolating cluster $P$. For the chain this function is trivially 1 for $p_A=p_B=1$ and 0 otherwise. $P$ is thus constant with a discontinuity at the percolation threshold, and $\beta=0$. 

In summary, for the one-dimensional case, the critical exponents for sublattice-selective percolation, specifying the behavior at the singular point $p_A=p_B=1$, are given by $\beta=0$, $\nu=1$, $\gamma=1$. These are the same  exponents as for conventional percolation in $d=1$~\cite{Stauffer2018}. Since only $p_A=p_B=1$ percolates, this conclusion may not be surprising. 

\subsection{Percolation threshold line for general $z$}

\begin{figure}[t]
     \centering
      \includegraphics[width=0.3\textwidth]{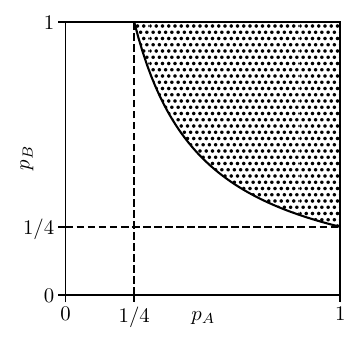}
     \caption{Phase diagram of sublattice-selective percolation on the Bethe lattice for $z=3$.}
     \label{Fig:Bethe_phasediag}
\end{figure}

The percolation condition on the Bethe lattice is that a chain of occupied sites reaches out to infinity. In other words, the system percolates if from some occupied site there exists a path of connected occupied sites that never ends (note that on the Bethe lattice such a path cannot form loops). If a chain of sites from a specific site up to some other site are occupied, at least 1 of the two neighbors going outwards also has to be occupied to have a percolating cluster. Hence,  the mean number of occupied sites going outwards has to be at least one in order to form a percolating cluster. Furthermore, on the Bethe lattice the $A$ and $B$ sublattices alternate. 
The mean number of occupied sites going outwards can thus be calculated for $z$ neighbors by using two nested binomial distributions, one  accounting for the mean number of $A$-sublattice sites and the other for the mean number of $B$-sublattice sites. This leads to the condition 
\begin{eqnarray}
    1\overset{!}{=}&&\sum_{k=0}^{z-1}\binom{z-1}{k}p_A^k(1-p_A)^{z-1-k} \nonumber \\
    && \times \sum_{l=0}^{(z-1)k}l\binom{(z-1)k}{l}p_B^l(1-p_B)^{(z-1)k-l},
\end{eqnarray}
which can be simplified to yield the condition for the percolation transition,
\begin{equation}\label{Eq:Bethe_PT_z}
p_A \: p_B= \frac{1}{(z-1)^2}.
\end{equation}
Note that this result is also correct for $z=2$, treated in the previous section. The percolation threshold line for $z=3$ is at $p_A\:p_B=1/4$, which is plotted in Fig.~\ref{Fig:Bethe_phasediag}. The larger $z$ is, the bigger the area of the phase diagram that is percolating, but here only  $z=3$ will be further examined. 

\subsection{Scaling exponents for $z=3$}

To access the critical exponents,  three state functions will again be calculated: The strength of the percolating cluster $P$, the mean cluster size $S$, and the correlation length $\xi$. Here, we focus on $z=3$. 

\begin{figure}[t]
     \centering
      \includegraphics[width=0.35\textwidth]{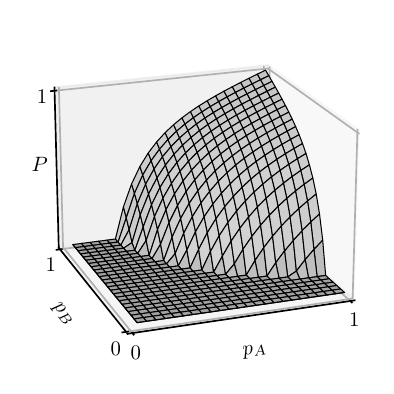}
     \caption{Strength $P$ of the percolating cluster for sublattice-selective percolation on the Bethe lattice for $z=3$.}
     \label{Fig:Bethe_P}
\end{figure}

\begin{figure}[t]
     \centering
      \includegraphics[width=0.5\textwidth]{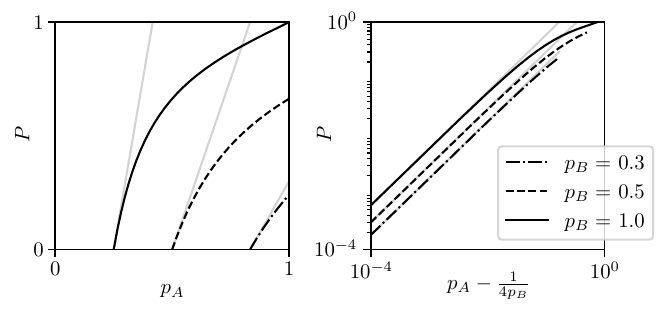}
     \caption{Strength $P$ of the percolating cluster as a function of $p_A$ for various values of $p_B$ (a) along with the function $6p_Ap_B-3/2$ and (b) in shifted log-log form near the percolation threshold.}
     \label{Fig:Bethe_P_scaling}
\end{figure}

To obtain $P$, we  first calculate the probability of a path not leading to infinity. Let $w_{A}$ ($w_B$) be the probability that an arbitrary $A$($B$) sublattice site does not  lead to infinity, respectively. An $A$ sublattice site does not lead to infinity if the site itself is not occupied (given by the probability $1-p_A$) or if the two neighbors leading outwards do not lead to infinity (given by the probability $p_A w_B^2$), and similarly for a $B$ sublattice site. Altogether, this gives the recursive expressions
\begin{align}
    w_A&=(1-p_A)+p_Aw_B^2,\label{eq:infd_QA}\\
    w_B&=(1-p_B)+p_Bw_A^2.\label{eq:infd_QB}
\end{align}
This nonlinear system of equations has polynomials of degree four as solutions. The trivial solution is $w_A=w_B=1$, while the other real solution, denoted by $w^P_A, w^P_B$ is rather lengthy, and given explicitly as a function of $p_A$ and $p_B$ in App.~\ref{App}. We note that for larger coordination number $z$, the corresponding recursion equations lead to higher-order equations for $w_A$ and $w_B$, for which closed expressions for the roots are known not not exist from Galois theory. For $p_A\: p_B<1/4$ there exists no infinite cluster, and thus the trivial solution applies, while for $p_A \: p_B\geq 1/4$ the other real solution is valid, i.e., 
\begin{equation}
    w_{A/B}=
    \begin{dcases}
        1, &4p_A\: p_B<1,\\
        w^P_{A/B}(p_A, p_B), &4p_A\: p_B\geq 1.
    \end{dcases}
\end{equation}
From these probabilities, the strength of the percolating cluster $P$ can be calculated as 
\begin{equation}
    P(p_A, p_B)=\frac12 \left(p_A(1-w_B^3)+p_B(1-w_A^3)\right),
\end{equation}
since the probability that the center site is from the $A$ or $B$ sublattice is $1/2$ and the probability that the site is occupied and at least one neighbor leads to infinity equals $p_{A/B}(1-w_{B/A}^3)$. The strength of the percolating cluster is plotted in Fig.~\ref{Fig:Bethe_P}.
In Fig.~\ref{Fig:Bethe_P_scaling}, cuts of $P$ at three different values of $p_B$ are shown, along with the function $6p_Ap_B-3/2$, which yields the leading asymptotic of $P$ for conventional site-percolation on the Bethe lattice~\cite{Stauffer2018}. We thus find that near the percolation threshold line, i.e., for $p_A \gtrsim 1/(4p_B), p_B\geq 1/4$, the scaling behavior
\begin{equation}
    P(p_A, p_B)\sim -3/2+6p_Ap_B=-6p_B\left(\frac{1}{4p_B}-p_A\right),
\end{equation}
emerges, implying a critical exponent $\beta=1$.

Next, the mean cluster size $S$ will be calculated. For this, an arbitrary occupied starting site is considered and the mean mass of the cluster it belongs to is calculated. This approach is only valid below the percolation threshold line, i.e., for $4p_Ap_B<1$, because otherwise the mass of the percolating cluster has to be accounted for as well. If the starting site belongs to  the $A$($B$) sublattice, then let $S_B$ ($S_A$) be the contribution to the mean cluster size from each neighbor. The mean cluster size then is composed of the contribution from the center site plus the contribution from the three neighbors, so that 
\begin{equation}
    S(p_A, p_B) = 1+3 \frac{1}{p_A+p_B}(p_AS_B+p_BS_A).
\end{equation}
The contributions $S_A$ and $S_B$ can be accessed recursively: If the starting site belongs to  the $A$ sublattice, then a given neighbor only contributes a non--zero $S_B$ with probability $p_B$. The size of this contribution is the site itself plus $S_A$ for the two neighbors going outwards. With a similar reasoning for the $B$ sublattice, we thus obtain 
\begin{equation}
    \begin{rcases}
        S_A=p_A(1+2S_B),\\
        S_B=p_B(1+2S_A),
    \end{rcases}
    \Rightarrow \quad
    \begin{array}{r@{\;}l}
        S_A=\frac{2p_Ap_B+p_B}{1-4p_Bp_A},\\
        S_B=\frac{2p_Ap_B+p_A}{1-4p_Bp_A}.
    \end{array}\label{eq:infd_T}
\end{equation}
Both $S_A$ and $S_B$ diverge upon approaching the threshold condition $p_A p_B =\frac{1}{4}$. Hence, the behavior of $S$ near the percolation threshold line can be described by
\begin{equation}
    S(p_A, p_B)\sim \left(\frac{1}{4p_B}-p_A\right)^{-1}, 
\end{equation}
giving the critical exponent $\gamma=1$.

Finally, the correlation length will be calculated. Here, the topological distance $r_{t}$ and Euclidean distance $r$ have to be distinguished. The topological distance between two sites,  also called the chemical distance, is defined as the number of bonds that connect them, in contrast to the Euclidean distance, which is the distance that these points are apart in space. For $d=\infty$, the conversion becomes easy because all bonds are pair--wise perpendicular, so that the Euclidean distance can be calculated by the generalized Pythagorean theorem as $r=\sqrt{r_{t}}$ ~\cite{Peruggi1984}. 

The correlation function $g(r)=g(\sqrt{r_{t}})$, for $4p_Ap_B<1$, equals  the mean number of occupied sites that are within the same cluster as a given occupied site, a topological distance $r_{t}$ apart -- corresponding to the Euclidean distance $r=\sqrt{r_{t}}$. At $r_{t}=0$ the center site is already occupied, while at $r_{t}=1$ the site has 3 neighbors and from there on the number of neighbors doubles in each step, and  the probabilities are otherwise the same as for the case of the chain, so that we obtain
\begin{equation}
    g(\sqrt{r_{t}})=
    \begin{dcases}
        1, &r_{t}=0\\
        3\cdot 2^{r_{t}-1} \frac{2}{p_A+p_B}p_A^\frac{(r_{t}+1)}{2}p_B^\frac{(r_{t}+1)}{2}, &\text{$r_{t}$ odd}\\
        3\cdot 2^{r_{t}-1} p_A^{r_{t}/2}p_B^{r_{t}/2}, &\text{$r_{t}$ even}
    \end{dcases}\nonumber
\end{equation}

The Euclidean correlation length can also be expressed in terms of the topological distance, since 
\begin{equation}
    \xi^2=\frac{\sum_{r\in \{0, 1, \sqrt{2}, \sqrt{3}, ...\}} r^2g(r)}{\sum_{r\in \{0, 1, \sqrt{2}, \sqrt{3}, ...\}} g(r)}
    =\frac{\sum_{r_{t}\in \mathbb{N}_0} r_{t}g(\sqrt{r_{t}})}{\sum_{r_{t}\in \mathbb{N}_0} g(\sqrt{r_{t}})}.\label{eq:infd_xi_def}
\end{equation}
We note that the identity $\sum_{r}g(r)=S$
is again satisfied, and hence the denominator of Eq.~(\ref{eq:infd_xi_def}) behaves like $(1-4p_Ap_B)^{-1}$ for $p_A\sim 1/{p_B}$. Furthermore, we find that the numerator behaves like $(1-4p_Ap_B)^{-2}$ for $p_A\sim {1}/{p_B}$, and together the singularity of $\xi$ near the percolation threshold is given by 
\begin{equation}
    \xi(p_A, p_B)\sim \left(\frac{1}{4p_B}-p_A\right)^{-\frac12}, 
\end{equation}
so that  $\nu=1/2$.
We thus obtain the same critical exponents as for conventional percolation on the Bethe lattice.

\section{Conclusions}\label{Sec:Conclusions}

We presented an adapted  Newman--Ziff algorithm to   examine sublattice-selective percolation on bipartite lattices and we applied it to accurately determine the percolation threshold lines for the square lattice and the planar Lieb lattice. The latter case relates to  mixed bond-site percolation on the square lattice, and our numerical results for the percolation threshold line refine previous estimates. Our numerical estimates for the critical exponents at the percolation transition are consistent with the universality class for percolation in $d=2$.

\begin{figure}[t]
     \centering
      \includegraphics[width=0.3\textwidth]{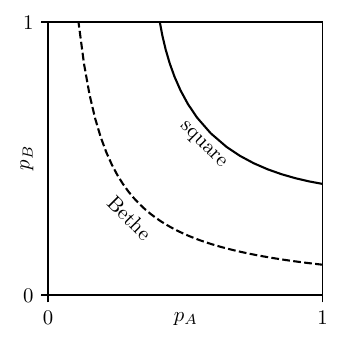}
     \caption{Comparison between the percolation threshold line for sublattice-selective percolation on the square lattice (solid line)  and the Bethe lattice with $z=4$ (dashed line).}
     \label{Fig:Bethe_sq_comp}
\end{figure}

In addition, we examined sublattice-selective percolation on the Bethe lattice, which was in fact the  geometry considered in the pioneering work on percolation by Flory in the context of polymerization~\cite{Flory1941}. For this case, we obtained the exact percolation threshold line and critical exponents for $z=2$ and $z=3$, which are again in accord with the values for conventional percolation. Given that the sites of the square lattice have a uniform coordination number of $z=4$, it may be interesting to compare its phase diagram to that of the Bethe lattice with $z=4$. Such a comparison is shown in Fig.~\ref{Fig:Bethe_sq_comp}, illustrating the substantially larger percolating regime for the latter case. Indeed, the absence of closed loops for a chain of connected sites leads to an overall prior appearance of the percolating cluster upon increasing the site occupation probabilities on the Bethe lattice. We checked explicitly, that a rescaling of the analytic expression for the percolation threshold line  for the $z=4$ Bethe lattice to match the ends points at $p_A=1$ and $p_B=1$ does not fit the full numerical data. 

Related to the last point, we note that the authors of Ref.~\cite{Tarasevich1999} considered various proposed analytic expressions for the general functional form of the percolation threshold line for mixed bond-site percolation. They observed clear systematic deviations from the numerical data for low-coordinated lattices. Based on our refined data on the Lieb lattice, we can also exclude the validity of these expressions for the square lattice (cf. Appendix~\ref{AppB}). It thus remains an interesting open question for further research to derive analytical expressions for the percolation threshold line for sublattice-selective as well as mixed bond-site percolation, respectively. We hope that our refined numerical data will be valuable as a benchmark for such investigations. Finally, let us note that sublattice-selective percolation can also be related to the problem of antisite defect percolation, such as that considered in Ref.~\cite{Tarasevich2003} for the case of a simple cubic lattice. Certainly, the adapted Newman-Ziff algorithm can be of use also for further investigations of such generalized percolation problems. 

\section*{Acknowledgements}

We thank Nils Caci for discussions. Furthermore, we acknowledge support by the Deutsche Forschungsgemeinschaft (DFG) through RTG 1995, and we thank the IT Center at RWTH Aachen University for access to computing time.

\appendix
\section{Non-trivial real solution for the $z=3$ Bethe lattice}\label{App}

The non-trivial real solution to Eqs.~(\ref{eq:infd_QA}) and (\ref{eq:infd_QB}), obtained using the Sympy Python library, is
\begin{widetext}
\begin{align*}
    w^P_{A/B} = &-\frac{1}{3}\left( \frac{1 - a_{A/B}}{c_{A/B}} + c_{A/B} + 1\right),\quad \mathrm{where}\\
    a_{A/B}=&\frac{3  \left(2 - p_{B/A}\right)}{p_{B/A}},\\
    b_{A/B}=&\frac{27 \left(- p_{A/B} p_{B/A}^{2} + 2 p_{A/B} p_{B/A} - 1\right)}{p_{A/B} p_{B/A}^{2}},\\
    c_{A/B}=&\sqrt[3]{\frac{\sqrt{- 4 \left(1 - a_{A/B}\right)^{3} + \left(2 - 3a_{A/B} + b_{A/B}\right)^{2}}}{2} + 1 - \frac{3}{2}a_{A/B} + b_{A/B}}.
\end{align*}
\end{widetext}

\section{Comparison for mixed bond-site percolation on the square lattice  }\label{AppB}
\begin{figure}[t]
     \centering
      \includegraphics[width=0.5\textwidth]{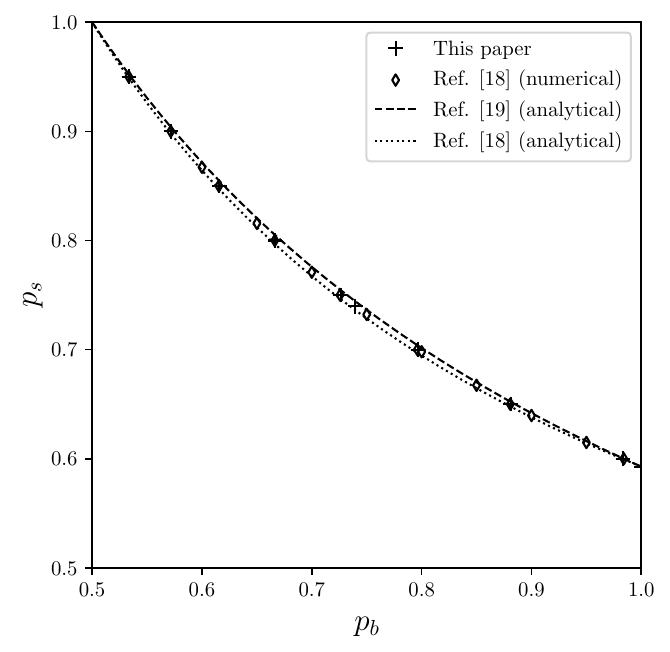}
     \caption{Comparison between the estimates for the percolation threshold line for mixed bond-site percolation on the square lattice as obtained from the adapted Newman-Ziff algorithm, the numerical estimates from Ref.~\cite{Tarasevich1999}, as well as the analytical formulas proposed in Ref.~\cite{Yanuka1990} (their Eq.~(3)), and Ref.~\cite{Tarasevich1999} (their Eq.~(19)), respectively.}
     \label{Fig:comp}
\end{figure}

In Fig.~\ref{Fig:comp}, we compare our results for the transition line for mixed bond-site percolation on the square lattice (as discussed in Sec.~\ref{Sec:Lieb}, this relates to sublattice-selective percolation on the Lieb lattice) to previous numerical estimates from Ref.~\cite{Tarasevich1999}, and two proposed analytical formulas for this transition line, taken from Ref.~\cite{Yanuka1990} (their Eq.~(3)), and 
Ref.~\cite{Tarasevich1999} (their Eq.~(19)), respectively. We find that our data, based on the adapted Newman-Ziff algorithm, are in accord with the numerical estimates reported in  Ref.~\cite{Tarasevich1999} and exclude the validity of both proposed analytical formulas.

\bibliography{main.bbl}

\begin{thebibliography}{25}%
\makeatletter
\providecommand \@ifxundefined [1]{%
 \@ifx{#1\undefined}
}%
\providecommand \@ifnum [1]{%
 \ifnum #1\expandafter \@firstoftwo
 \else \expandafter \@secondoftwo
 \fi
}%
\providecommand \@ifx [1]{%
 \ifx #1\expandafter \@firstoftwo
 \else \expandafter \@secondoftwo
 \fi
}%
\providecommand \natexlab [1]{#1}%
\providecommand \enquote  [1]{``#1''}%
\providecommand \bibnamefont  [1]{#1}%
\providecommand \bibfnamefont [1]{#1}%
\providecommand \citenamefont [1]{#1}%
\providecommand \href@noop [0]{\@secondoftwo}%
\providecommand \href [0]{\begingroup \@sanitize@url \@href}%
\providecommand \@href[1]{\@@startlink{#1}\@@href}%
\providecommand \@@href[1]{\endgroup#1\@@endlink}%
\providecommand \@sanitize@url [0]{\catcode `\\12\catcode `\$12\catcode
  `\&12\catcode `\#12\catcode `\^12\catcode `\_12\catcode `\%12\relax}%
\providecommand \@@startlink[1]{}%
\providecommand \@@endlink[0]{}%
\providecommand \url  [0]{\begingroup\@sanitize@url \@url }%
\providecommand \@url [1]{\endgroup\@href {#1}{\urlprefix }}%
\providecommand \urlprefix  [0]{URL }%
\providecommand \Eprint [0]{\href }%
\providecommand \doibase [0]{http://dx.doi.org/}%
\providecommand \selectlanguage [0]{\@gobble}%
\providecommand \bibinfo  [0]{\@secondoftwo}%
\providecommand \bibfield  [0]{\@secondoftwo}%
\providecommand \translation [1]{[#1]}%
\providecommand \BibitemOpen [0]{}%
\providecommand \bibitemStop [0]{}%
\providecommand \bibitemNoStop [0]{.\EOS\space}%
\providecommand \EOS [0]{\spacefactor3000\relax}%
\providecommand \BibitemShut  [1]{\csname bibitem#1\endcsname}%
\let\auto@bib@innerbib\@empty
\bibitem [{\citenamefont {Flory}(1941)}]{Flory1941}%
  \BibitemOpen
  \bibfield  {author} {\bibinfo {author} {\bibfnamefont {P.~J.}\ \bibnamefont
  {Flory}},\ }\href@noop {} {\bibfield  {journal} {\bibinfo  {journal} {Journal
  of the American Chemical Society}\ }\textbf {\bibinfo {volume} {63}}
  (\bibinfo {year} {1941})}\BibitemShut {NoStop}%
\bibitem [{\citenamefont {Stauffer}\ and\ \citenamefont
  {Aharony}(2018)}]{Stauffer2018}%
  \BibitemOpen
  \bibfield  {author} {\bibinfo {author} {\bibfnamefont {D.}~\bibnamefont
  {Stauffer}}\ and\ \bibinfo {author} {\bibfnamefont {A.}~\bibnamefont
  {Aharony}},\ }\href {\doibase 10.1201/9781315274386} {\emph {\bibinfo {title}
  {{Introduction To Percolation Theory}}}}\ (\bibinfo  {publisher} {Taylor \&
  Francis},\ \bibinfo {year} {2018})\BibitemShut {NoStop}%
\bibitem [{\citenamefont {Lee}(2008)}]{Lee2008}%
  \BibitemOpen
  \bibfield  {author} {\bibinfo {author} {\bibfnamefont {M.~J.}\ \bibnamefont
  {Lee}},\ }\href {\doibase 10.1103/PhysRevE.78.031131} {\bibfield  {journal}
  {\bibinfo  {journal} {Physical Review E - Statistical, Nonlinear, and Soft
  Matter Physics}\ }\textbf {\bibinfo {volume} {78}} (\bibinfo {year} {2008}),\
  10.1103/PhysRevE.78.031131}\BibitemShut {NoStop}%
\bibitem [{\citenamefont {Jacobsen}(2015)}]{Jacobsen2015}%
  \BibitemOpen
  \bibfield  {author} {\bibinfo {author} {\bibfnamefont {J.~L.}\ \bibnamefont
  {Jacobsen}},\ }\href {\doibase 10.1088/1751-8113/48/45/454003} {\bibfield
  {journal} {\bibinfo  {journal} {Journal of Physics A: Mathematical and
  Theoretical}\ }\textbf {\bibinfo {volume} {48}},\ \bibinfo {pages} {454003}
  (\bibinfo {year} {2015})}\BibitemShut {NoStop}%
\bibitem [{\citenamefont {Smirnov}\ and\ \citenamefont
  {Werner}(2001)}]{Smirnov2001}%
  \BibitemOpen
  \bibfield  {author} {\bibinfo {author} {\bibfnamefont {S.}~\bibnamefont
  {Smirnov}}\ and\ \bibinfo {author} {\bibfnamefont {W.}~\bibnamefont
  {Werner}},\ }\href {https://api.semanticscholar.org/CorpusID:6837772}
  {\bibfield  {journal} {\bibinfo  {journal} {Mathematical Research Letters}\
  }\textbf {\bibinfo {volume} {8}},\ \bibinfo {pages} {729} (\bibinfo {year}
  {2001})}\BibitemShut {NoStop}%
\bibitem [{\citenamefont {Scholl}\ and\ \citenamefont
  {Binder}(1980)}]{Scholl1980}%
  \BibitemOpen
  \bibfield  {author} {\bibinfo {author} {\bibfnamefont {F.}~\bibnamefont
  {Scholl}}\ and\ \bibinfo {author} {\bibfnamefont {K.}~\bibnamefont
  {Binder}},\ }\href@noop {} {\bibfield  {journal} {\bibinfo  {journal}
  {Zeitschrift für Physik B Condensed Matter}\ }\textbf {\bibinfo {volume}
  {39}} (\bibinfo {year} {1980})}\BibitemShut {NoStop}%
\bibitem [{\citenamefont {Sykes}\ and\ \citenamefont
  {Essam}(1964)}]{Sykes1964}%
  \BibitemOpen
  \bibfield  {author} {\bibinfo {author} {\bibfnamefont {M.~F.}\ \bibnamefont
  {Sykes}}\ and\ \bibinfo {author} {\bibfnamefont {J.~W.}\ \bibnamefont
  {Essam}},\ }\href@noop {} {\bibfield  {journal} {\bibinfo  {journal} {Journal
  of Mathematical Physics}\ }\textbf {\bibinfo {volume} {5}} (\bibinfo {year}
  {1964})}\BibitemShut {NoStop}%
\bibitem [{\citenamefont {Idogaki}\ and\ \citenamefont
  {Uryû}(1982)}]{Idogaki1982a}%
  \BibitemOpen
  \bibfield  {author} {\bibinfo {author} {\bibfnamefont {T.}~\bibnamefont
  {Idogaki}}\ and\ \bibinfo {author} {\bibfnamefont {N.}~\bibnamefont
  {Uryû}},\ }\href {\doibase https://doi.org/10.1016/0375-9601(82)90633-8}
  {\bibfield  {journal} {\bibinfo  {journal} {Physics Letters A}\ }\textbf
  {\bibinfo {volume} {90}},\ \bibinfo {pages} {367} (\bibinfo {year}
  {1982})}\BibitemShut {NoStop}%
\bibitem [{\citenamefont {Idogaki}\ and\ \citenamefont
  {Uryu}(1982)}]{Idogaki1982b}%
  \BibitemOpen
  \bibfield  {author} {\bibinfo {author} {\bibfnamefont {T.}~\bibnamefont
  {Idogaki}}\ and\ \bibinfo {author} {\bibfnamefont {N.}~\bibnamefont {Uryu}},\
  }\href {\doibase 10.1088/0022-3719/15/31/001} {\bibfield  {journal} {\bibinfo
   {journal} {Journal of Physics C: Solid State Physics}\ }\textbf {\bibinfo
  {volume} {15}},\ \bibinfo {pages} {L1077} (\bibinfo {year}
  {1982})}\BibitemShut {NoStop}%
\bibitem [{\citenamefont {Idogaki}\ and\ \citenamefont
  {Uryû}(1983)}]{Idogaki1983a}%
  \BibitemOpen
  \bibfield  {author} {\bibinfo {author} {\bibfnamefont {T.}~\bibnamefont
  {Idogaki}}\ and\ \bibinfo {author} {\bibfnamefont {N.}~\bibnamefont
  {Uryû}},\ }\href {\doibase https://doi.org/10.1016/0304-8853(83)90888-0}
  {\bibfield  {journal} {\bibinfo  {journal} {Journal of Magnetism and Magnetic
  Materials}\ }\textbf {\bibinfo {volume} {31-34}},\ \bibinfo {pages} {1257}
  (\bibinfo {year} {1983})}\BibitemShut {NoStop}%
\bibitem [{\citenamefont {Idogaki}\ \emph {et~al.}(1995)\citenamefont
  {Idogaki}, \citenamefont {Hitaka},\ and\ \citenamefont
  {Masuda}}]{Idogaki1995}%
  \BibitemOpen
  \bibfield  {author} {\bibinfo {author} {\bibfnamefont {T.}~\bibnamefont
  {Idogaki}}, \bibinfo {author} {\bibfnamefont {M.}~\bibnamefont {Hitaka}}, \
  and\ \bibinfo {author} {\bibfnamefont {K.}~\bibnamefont {Masuda}},\ }\href
  {\doibase https://doi.org/10.1016/0304-8853(94)00507-9} {\bibfield  {journal}
  {\bibinfo  {journal} {Journal of Magnetism and Magnetic Materials}\ }\textbf
  {\bibinfo {volume} {140-144}},\ \bibinfo {pages} {1517} (\bibinfo {year}
  {1995})}\BibitemShut {NoStop}%
\bibitem [{\citenamefont {Ueda}\ \emph {et~al.}(2000)\citenamefont {Ueda},
  \citenamefont {Tanaka}, \citenamefont {Iwashita},\ and\ \citenamefont
  {Idogaki}}]{Ueda2000}%
  \BibitemOpen
  \bibfield  {author} {\bibinfo {author} {\bibfnamefont {H.}~\bibnamefont
  {Ueda}}, \bibinfo {author} {\bibfnamefont {A.}~\bibnamefont {Tanaka}},
  \bibinfo {author} {\bibfnamefont {T.}~\bibnamefont {Iwashita}}, \ and\
  \bibinfo {author} {\bibfnamefont {T.}~\bibnamefont {Idogaki}},\ }\href
  {\doibase https://doi.org/10.1016/S0921-4526(99)02657-5} {\bibfield
  {journal} {\bibinfo  {journal} {Physica B: Condensed Matter}\ }\textbf
  {\bibinfo {volume} {284-288}},\ \bibinfo {pages} {1201} (\bibinfo {year}
  {2000})}\BibitemShut {NoStop}%
\bibitem [{\citenamefont {Newman}\ and\ \citenamefont
  {Ziff}(2000)}]{Newman2000}%
  \BibitemOpen
  \bibfield  {author} {\bibinfo {author} {\bibfnamefont {M.~E.~J.}\
  \bibnamefont {Newman}}\ and\ \bibinfo {author} {\bibfnamefont {R.~M.}\
  \bibnamefont {Ziff}},\ }\href {\doibase 10.1103/PhysRevLett.85.4104}
  {\bibfield  {journal} {\bibinfo  {journal} {Phys. Rev. Lett.}\ }\textbf
  {\bibinfo {volume} {85}},\ \bibinfo {pages} {4104} (\bibinfo {year}
  {2000})}\BibitemShut {NoStop}%
\bibitem [{\citenamefont {Newman}\ and\ \citenamefont
  {Ziff}(2001)}]{Newman2001}%
  \BibitemOpen
  \bibfield  {author} {\bibinfo {author} {\bibfnamefont {M.~E.~J.}\
  \bibnamefont {Newman}}\ and\ \bibinfo {author} {\bibfnamefont {R.~M.}\
  \bibnamefont {Ziff}},\ }\href {\doibase 10.1103/PhysRevE.64.016706}
  {\bibfield  {journal} {\bibinfo  {journal} {Phys. Rev. E}\ }\textbf {\bibinfo
  {volume} {64}},\ \bibinfo {pages} {016706} (\bibinfo {year}
  {2001})}\BibitemShut {NoStop}%
\bibitem [{\citenamefont {Rosales~Herrera}\ \emph {et~al.}(2021)\citenamefont
  {Rosales~Herrera}, \citenamefont {Ramírez}, \citenamefont {Martínez},
  \citenamefont {Cruz-Suárez}, \citenamefont {Fernández~Téllez},
  \citenamefont {López-Olguín},\ and\ \citenamefont
  {Aragón~García}}]{rosales_herrera_percolation-intercropping_2021}%
  \BibitemOpen
  \bibfield  {author} {\bibinfo {author} {\bibfnamefont {D.}~\bibnamefont
  {Rosales~Herrera}}, \bibinfo {author} {\bibfnamefont {J.~E.}\ \bibnamefont
  {Ramírez}}, \bibinfo {author} {\bibfnamefont {M.~I.}\ \bibnamefont   
  {Martínez}}, \bibinfo {author} {\bibfnamefont {H.}~\bibnamefont
  {Cruz-Suárez}}, \bibinfo {author} {\bibfnamefont {A.}~\bibnamefont
  {Fernández~Téllez}}, \bibinfo {author} {\bibfnamefont {J.~F.}\ \bibnamefont
  {López-Olguín}}, \ and\ \bibinfo {author} {\bibfnamefont {A.}~\bibnamefont
  {Aragón~García}},\ }\href {\doibase 10.1063/5.0044714} {\bibfield
  {journal} {\bibinfo  {journal} {Chaos: An Interdisciplinary Journal of
  Nonlinear Science}\ }\textbf {\bibinfo {volume} {31}},\ \bibinfo {pages}
  {063105} (\bibinfo {year} {2021})}\BibitemShut {NoStop}%
\bibitem [{\citenamefont {Oliveira}\ \emph {et~al.}(2021)\citenamefont
  {Oliveira}, \citenamefont {de~Lima}, \citenamefont {Costa},\ and\
  \citenamefont {dos Santos}}]{Oliveria2021}%
  \BibitemOpen
  \bibfield  {author} {\bibinfo {author} {\bibfnamefont {W.~S.}\ \bibnamefont
  {Oliveira}}, \bibinfo {author} {\bibfnamefont {J.~P.}\ \bibnamefont
  {de~Lima}}, \bibinfo {author} {\bibfnamefont {N.~C.}\ \bibnamefont {Costa}},
  \ and\ \bibinfo {author} {\bibfnamefont {R.~R.}\ \bibnamefont {dos Santos}},\
  }\href {\doibase 10.1103/PhysRevE.104.064122} {\bibfield  {journal} {\bibinfo
   {journal} {Phys. Rev. E}\ }\textbf {\bibinfo {volume} {104}},\ \bibinfo
  {pages} {064122} (\bibinfo {year} {2021})}\BibitemShut {NoStop}%
\bibitem [{\citenamefont {Hoshen}\ and\ \citenamefont
  {Kopelman}(1976)}]{Hoshen1976}%
  \BibitemOpen
  \bibfield  {author} {\bibinfo {author} {\bibfnamefont {J.}~\bibnamefont
  {Hoshen}}\ and\ \bibinfo {author} {\bibfnamefont {R.}~\bibnamefont
  {Kopelman}},\ }\href@noop {} {\bibfield  {journal} {\bibinfo  {journal}
  {Phys. Rev. B}\ }\textbf {\bibinfo {volume} {14}} (\bibinfo {year}
  {1976})}\BibitemShut {NoStop}%
\bibitem [{\citenamefont {Kesten}(1980)}]{Kesten1980}%
  \BibitemOpen
  \bibfield  {author} {\bibinfo {author} {\bibfnamefont {H.}~\bibnamefont
  {Kesten}},\ }\href {https://api.semanticscholar.org/CorpusID:3143683}
  {\bibfield  {journal} {\bibinfo  {journal} {Communications in Mathematical
  Physics}\ }\textbf {\bibinfo {volume} {74}},\ \bibinfo {pages} {41} (\bibinfo
  {year} {1980})}\BibitemShut {NoStop}%
\bibitem [{\citenamefont {Tarasevich}\ and\ \citenamefont {Van~der
  Mark}(1999)}]{Tarasevich1999}%
  \BibitemOpen
  \bibfield  {author} {\bibinfo {author} {\bibfnamefont {Y.~Y.}\ \bibnamefont
  {Tarasevich}}\ and\ \bibinfo {author} {\bibfnamefont {S.~C.}\ \bibnamefont
  {Van~der Mark}},\ }\href {\doibase 10.1142/S0129183199000978} {\bibfield
  {journal} {\bibinfo  {journal} {International Journal of Modern Physics C}\
  }\textbf {\bibinfo {volume} {10}},\ \bibinfo {pages} {1193} (\bibinfo {year}
  {1999})}\BibitemShut {NoStop}%
\bibitem [{\citenamefont {Yanuka}\ and\ \citenamefont
  {Englman}(1990)}]{Yanuka1990}%
  \BibitemOpen
  \bibfield  {author} {\bibinfo {author} {\bibfnamefont {M.}~\bibnamefont
  {Yanuka}}\ and\ \bibinfo {author} {\bibfnamefont {R.}~\bibnamefont
  {Englman}},\ }\href {\doibase 10.1088/0305-4470/23/7/010} {\bibfield
  {journal} {\bibinfo  {journal} {Journal of Physics A: Mathematical and
  General}\ }\textbf {\bibinfo {volume} {23}},\ \bibinfo {pages} {L339}
  (\bibinfo {year} {1990})}\BibitemShut {NoStop}%
\bibitem [{\citenamefont {González}\ \emph {et~al.}(2013)\citenamefont
  {González}, \citenamefont {Centres}, \citenamefont {Lebrecht}, \citenamefont
  {Ramirez-Pastor},\ and\ \citenamefont {Nieto}}]{Gonzalez2013}%
  \BibitemOpen
  \bibfield  {author} {\bibinfo {author} {\bibfnamefont {M.}~\bibnamefont
  {González}}, \bibinfo {author} {\bibfnamefont {P.}~\bibnamefont {Centres}},
  \bibinfo {author} {\bibfnamefont {W.}~\bibnamefont {Lebrecht}}, \bibinfo
  {author} {\bibfnamefont {A.}~\bibnamefont {Ramirez-Pastor}}, \ and\ \bibinfo
  {author} {\bibfnamefont {F.}~\bibnamefont {Nieto}},\ }\href {\doibase
  https://doi.org/10.1016/j.physa.2013.09.001} {\bibfield  {journal} {\bibinfo
  {journal} {Physica A: Statistical Mechanics and its Applications}\ }\textbf
  {\bibinfo {volume} {392}},\ \bibinfo {pages} {6330} (\bibinfo {year}
  {2013})}\BibitemShut {NoStop}%
\bibitem [{\citenamefont {Gonz\'alez-Flores}\ \emph {et~al.}(2021)\citenamefont
  {Gonz\'alez-Flores}, \citenamefont {Torres}, \citenamefont {Lebrecht},\ and\
  \citenamefont {Ramirez-Pastor}}]{Gonzalez2021}%
  \BibitemOpen
  \bibfield  {author} {\bibinfo {author} {\bibfnamefont {M.~I.}\ \bibnamefont
  {Gonz\'alez-Flores}}, \bibinfo {author} {\bibfnamefont {A.~A.}\ \bibnamefont
  {Torres}}, \bibinfo {author} {\bibfnamefont {W.}~\bibnamefont {Lebrecht}}, \
  and\ \bibinfo {author} {\bibfnamefont {A.~J.}\ \bibnamefont
  {Ramirez-Pastor}},\ }\href {\doibase 10.1103/PhysRevE.104.014130} {\bibfield
  {journal} {\bibinfo  {journal} {Phys. Rev. E}\ }\textbf {\bibinfo {volume}
  {104}},\ \bibinfo {pages} {014130} (\bibinfo {year} {2021})}\BibitemShut
  {NoStop}%
\bibitem [{\citenamefont {Torres}\ \emph {et~al.}(2022)\citenamefont {Torres},
  \citenamefont {González-Flores}, \citenamefont {Lebrecht},\ and\
  \citenamefont {Ramirez-Pastor}}]{Torres2022}%
  \BibitemOpen
  \bibfield  {author} {\bibinfo {author} {\bibfnamefont {A.}~\bibnamefont
  {Torres}}, \bibinfo {author} {\bibfnamefont {M.}~\bibnamefont
  {González-Flores}}, \bibinfo {author} {\bibfnamefont {W.}~\bibnamefont
  {Lebrecht}}, \ and\ \bibinfo {author} {\bibfnamefont {A.}~\bibnamefont
  {Ramirez-Pastor}},\ }\href {\doibase
  https://doi.org/10.1016/j.physa.2022.127897} {\bibfield  {journal} {\bibinfo
  {journal} {Physica A: Statistical Mechanics and its Applications}\ }\textbf
  {\bibinfo {volume} {604}},\ \bibinfo {pages} {127897} (\bibinfo {year}
  {2022})}\BibitemShut {NoStop}%
\bibitem [{\citenamefont {Peruggi}\ \emph {et~al.}(1984)\citenamefont
  {Peruggi}, \citenamefont {di~Liberto},\ and\ \citenamefont
  {Monroy}}]{Peruggi1984}%
  \BibitemOpen
  \bibfield  {author} {\bibinfo {author} {\bibfnamefont {F.}~\bibnamefont
  {Peruggi}}, \bibinfo {author} {\bibfnamefont {F.}~\bibnamefont {di~Liberto}},
  \ and\ \bibinfo {author} {\bibfnamefont {G.}~\bibnamefont {Monroy}},\ }\href
  {\doibase 10.1016/0378-4371(84)90110-9} {\bibfield  {journal} {\bibinfo
  {journal} {Physica A: Statistical Mechanics and its Applications}\ }\textbf
  {\bibinfo {volume} {123}} (\bibinfo {year} {1984}),\
  10.1016/0378-4371(84)90110-9}\BibitemShut {NoStop}%
\bibitem [{\citenamefont {Tarasevich}\ and\ \citenamefont
  {Manzhosova}(2003)}]{Tarasevich2003}%
  \BibitemOpen
  \bibfield  {author} {\bibinfo {author} {\bibfnamefont {Y.~Y.}\ \bibnamefont
  {Tarasevich}}\ and\ \bibinfo {author} {\bibfnamefont {E.~N.}\ \bibnamefont
  {Manzhosova}},\ }\href {\doibase 10.1142/S0129183103005480} {\bibfield
  {journal} {\bibinfo  {journal} {International Journal of Modern Physics C}\
  }\textbf {\bibinfo {volume} {14}},\ \bibinfo {pages} {1405} (\bibinfo {year}
  {2003})}\BibitemShut {NoStop}%
\end{thebibliography}%
\end{document}